\documentclass[a4paper,num-refs]{oup-contemporary}

\journal{gigascience}

\usepackage{graphicx}
\usepackage{siunitx}
\usepackage{subfigure}

\title{Hyperspectral Calibration of Art: Acquisition and Calibration Workflows}

\author[1,\authfn{1}]{Ruven Pillay}
\author[1]{Jon Y Hardeberg}
\author[1]{Sony George}

\affil[1]{NTNU - Norwegian University of Science and Technology, Department of Computer Science, Gj{\o}vik, Norway}
\affil[2]{C2RMF - Centre de Restauration et Recherche des Mus{\'e}es de France, Palais du Louvre, Paris, France}

\authnote{\authfn{1}Corresponding author: ruven.pillay@culture.gouv.fr}

\papercat{Preprint}

\runningauthor{Hyperspectral Calibration of Art: Acquisition and Calibration Workflows}

\jvolume{58}
\jnumber{1}
\jyear{2019}

\usepackage{natbib}
\bibpunct[, ]{(}{)}{;}{a}{}{,}

\begin{document}

\begin{frontmatter}
\maketitle
\begin{abstract}

Hyperspectral imaging has become an increasingly used tool in the analysis of works of art. However, the quality of the acquired data and the processing of that data to produce accurate and reproducible spectral image cubes can be a challenge to many cultural heritage users. The calibration of data that is both spectrally and spatially accurate is an essential step in order to obtain useful and relevant results from hyperspectral imaging. Data that is too noisy or inaccurate will produce sub-optimal results when used for pigment mapping, the detection of hidden features, change detection or for quantitative spectral documentation. To help address this, therefore, we will examine the specific acquisition and calibration workflows necessary for works of art. These workflows includes the key parameters that must be addressed during acquisition and the essential steps and issues at each of the stages required during post-processing in order to fully calibrate hyperspectral data. In addition we will look in detail at the key issues that affect data quality and propose practical solutions that can make significant differences to overall hyperspectral image quality.


\end{abstract}

\end{frontmatter}

\section{Introduction}

\subsection{Hyperspectral Imaging for Cultural Heritage}

Spectral reflectance imaging can be a valuable tool for analyzing and documenting works of art due to its ability to simultaneously capture both accurate spectral and spatial data. The introduction of spectral reflectance imaging technologies to the cultural heritage field occurred initially through multispectral imaging systems able to capture a handful of spectral bands \citep{saunders_image_1993,martinez_ten_2002,lahanier_crisatel_2002}. However, the development and commercial availability of push-broom hyperspectral cameras able to capture hundreds of narrow contiguous wavelengths soon became a practical possibility and a small number of institutions were able to develop bespoke equipment \citep{bacci_study_2005,delaney_visible_2010}.

\begin{figure*}[ht]
\centering
\subfigure[Horizontally-mounted push-broom hyperspectral system with motorized Y-axis scanning a manuscript.]{%
  \label{fig:equipment_horizontal}
  \resizebox*{8.0cm}{!}{\includegraphics{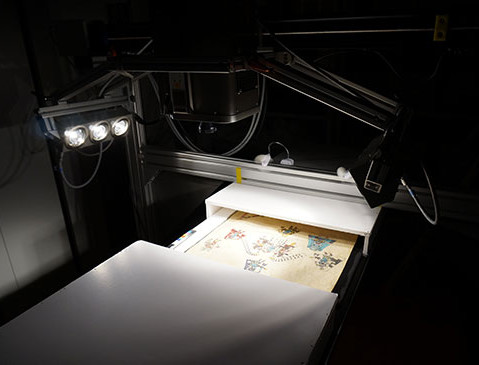}}}
  \hspace{5pt}
\subfigure[Vertically-mounted push-broom hyperspectral system with motorized XY axes scanning a painting.]{%
  \label{fig:equipment_vertical} %
  \resizebox*{5.0cm}{!}{\includegraphics{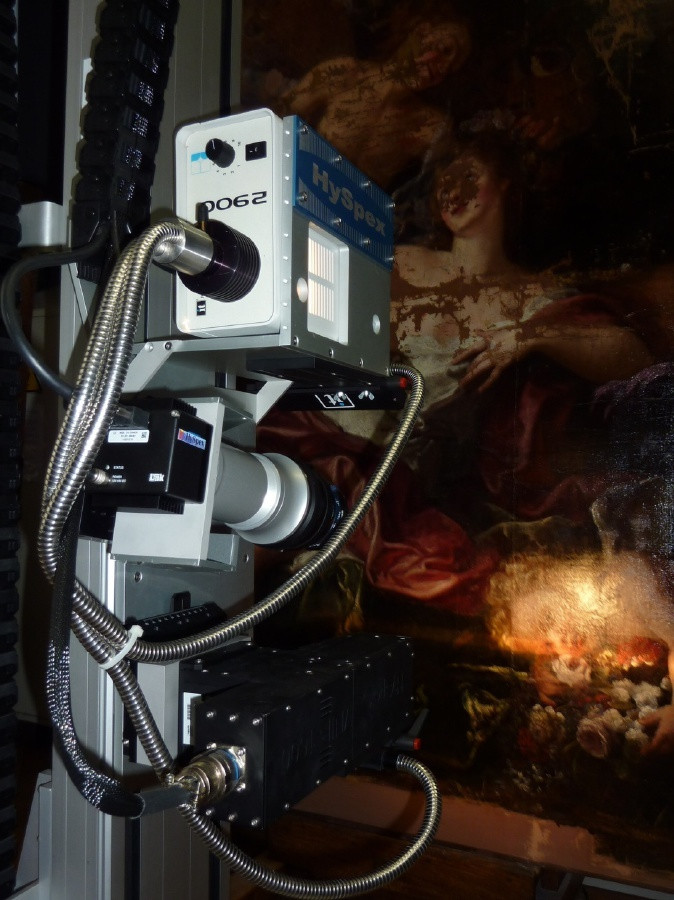}}}
\caption{Hyperspectral imaging equipment used in the cultural heritage field.}
\label{fig:equipment}
\end{figure*}

A number of system designs have been implemented for push-broom hyperspectral imaging of art. Standard laboratory systems for industrial use are usually horizontally mounted with the target lying flat on a horizontal surface. This set-up works well for manuscripts (figure \ref{fig:equipment_horizontal}), books or drawings. However, for easel paintings, a vertical alignment is better suited, which therefore requires a custom-made set-up, such as that in figure \ref{fig:equipment_vertical}. Various push-broom based hyperspectral imaging systems have been put in place for cultural heritage to analyze paintings \citep{bacci_study_2005, delaney_visible_2009}, codices \citep{snijders_using_2016, pottier_study_2017}, wall paintings \citep{liang_remote_2014} and many others. In almost all of these systems, the push-broom hyperspectral camera moves linearly along one or more axes, while the painting remains mounted on a static easel. It is also possible, however, to mount the hyperspectral camera on a motorized rotating tripod \citep{george_estimation_2016} or use a rotating scan mirror in front of the hyperspectral camera to enable snapshot imaging of the work of art \citep{delaney_visible_2010}.

Although the use of such equipment within cultural heritage institutions remains relatively rare, an increasing number of studies have made use of such equipment and hyperspectral imaging has successfully been used for a variety of applications including the revealing of otherwise hidden features and areas of lost material \citep{bacci_study_2005}, for materials identification and pigment mapping \citep{delaney_use_2014,deborah_pigment_2014} where methods and algorithms from the field of remote sensing have been applied as well as for documentation and visualization \citep{pillay_hyperspectral_2013}.

Nevertheless, a number of important challenges exist, which will need to be addressed before hyperspectral imaging can become more widely used within the cultural heritage sector. The sheer volumes of hyperspectral data that can be produced from an acquisition requires high-end processing, data storage capacity and the expertise to go with it. Such facilities are rarely available in museum environments, making this a major challenge. In addition, existing methodologies, algorithms and software for calibration, spectral mapping and unmixing methods etc. have been developed and optimized mainly for remote sensing applications. These will require careful customization and tuning to the needs of cultural heritage if the full potential of hyperspectral imaging is to be realized.

\subsection{Calibration Workflow}

The accurate calibration of spectral data is an essential step in order to obtain useful and relevant results from hyperspectral imaging. Colorimetric or spectral analysis of the data is highly sensitive to incorrect or sub-optimal calibration. The issue of spectral calibration has been well studied in the field of remote sensing \citep{woodward_hyperspectral_2009, eismann_hyperspectral_2012}, environmental imaging \citep{aspinall_considerations_2002}, plant biology \citep{behmann_calibration_2015} and for astronomy \citep{filacchione_calibration_2009}. For laboratory-based systems, there have been several studies carried out, in particular \citep{polder_calibration_2003, geladi_hyperspectral_2004, burger_hyperspectral_2005, burger_hyperspectral_2006, boldrini_hyperspectral_2012-1} and more recently \citep{khan_hytexila_2018}. However, none of these have addressed the specific requirements for the spectral imaging of works of art such as paintings or manuscripts.

In the following sections, we will, therefore, look at the issues arising from the imaging of works of art where the goal is to obtain high-resolution spectral as well as spatial data. We will consider the overall acquisition and calibration pipeline and examine ways to improve the accuracy and precision of the resulting data.

We will draw, in particular, on the results obtained from a round-robin test (RRT) of spectral imaging equipment \citep{george_study_2018}, which involved a number of institutions and a variety of different hyperspectral systems and set-ups. The RRT provided an important insight into how such equipment is used in real-world settings, the kind of data that is typically obtained and the specific problems and errors that can arise when applying hyperspectral imaging to works of art.

The results revealed a number of issues related to system calibration, reproducability and the lack of standardized workflows that have acted as a barrier to the greater take-up and use of hyperspectral imaging within the cultural heritage community.  We will, therefore, attempt to address some of the key issues by examining in detail the acquisition and processing steps, the factors that affect performance and common errors that can arise when using hyperspectral imaging and, thereby, help to provide a more standardized workflow and guidance for users seeking to make better use of their equipment.

A full workflow for the set-up and processing of hyperspectral acquisition for 2D artworks is described in the following sections.
The typical pipeline in spectral calibration involves first the removal of fixed pattern noise (dark current), pixel sensitivity normalization, denoising, illumination correction and finally spectral calibration to reflectance factor. Spatial geometric correction then needs to be carried out and finally mosaic assembly is required if the data has been acquired in multiple sections.

The resultant output from this calibration workflow is a spectral image cube that has been fully calibrated both radiometrically and spatially and which contains quantitative reflectance data suitable for use in applications such as materials mapping.

\section{Hyperspectral Acquisition Parameters}

Hyperspectral imaging systems are very sensitive to acquisition parameters and operating conditions. Ensuring the best possible acquisition conditions will not only help improve the quality of the resulting data but can also greatly simplify post-processing and data handling.



\subsection{Reference Targets}

\subsubsection{Reflectance Targets}

In order to be able to obtain quantitative spectral data, a known reference must be used to normalize the acquired spectra to absolute reflectance factor. To do this, an acquisition must be made of a known reflectance standard or set of reflectance standards under the same operating conditions and acquisition parameters used for the acquisition of the work of art itself. These conditions include lighting levels, integration time, scan speed, acquisition geometry, camera gain settings etc.

Ideal reference targets are uniform and diffuse lambertian planar surfaces that have constant reflectance across a wide spectral range. These are typically durable and chemically inert PTFE (Polytetrafluoroethylene) targets, the most widely used of which is Spectralon\textsuperscript{\textregistered} \footnote{manufactured by Labsphere Inc.}. Typically, a single reflectance standard is used and, indeed, the majority of the participants of the RRT acquired a single 99\% reference standard image.

\begin{figure}[h]
\centering
\includegraphics[width=0.4\textwidth]{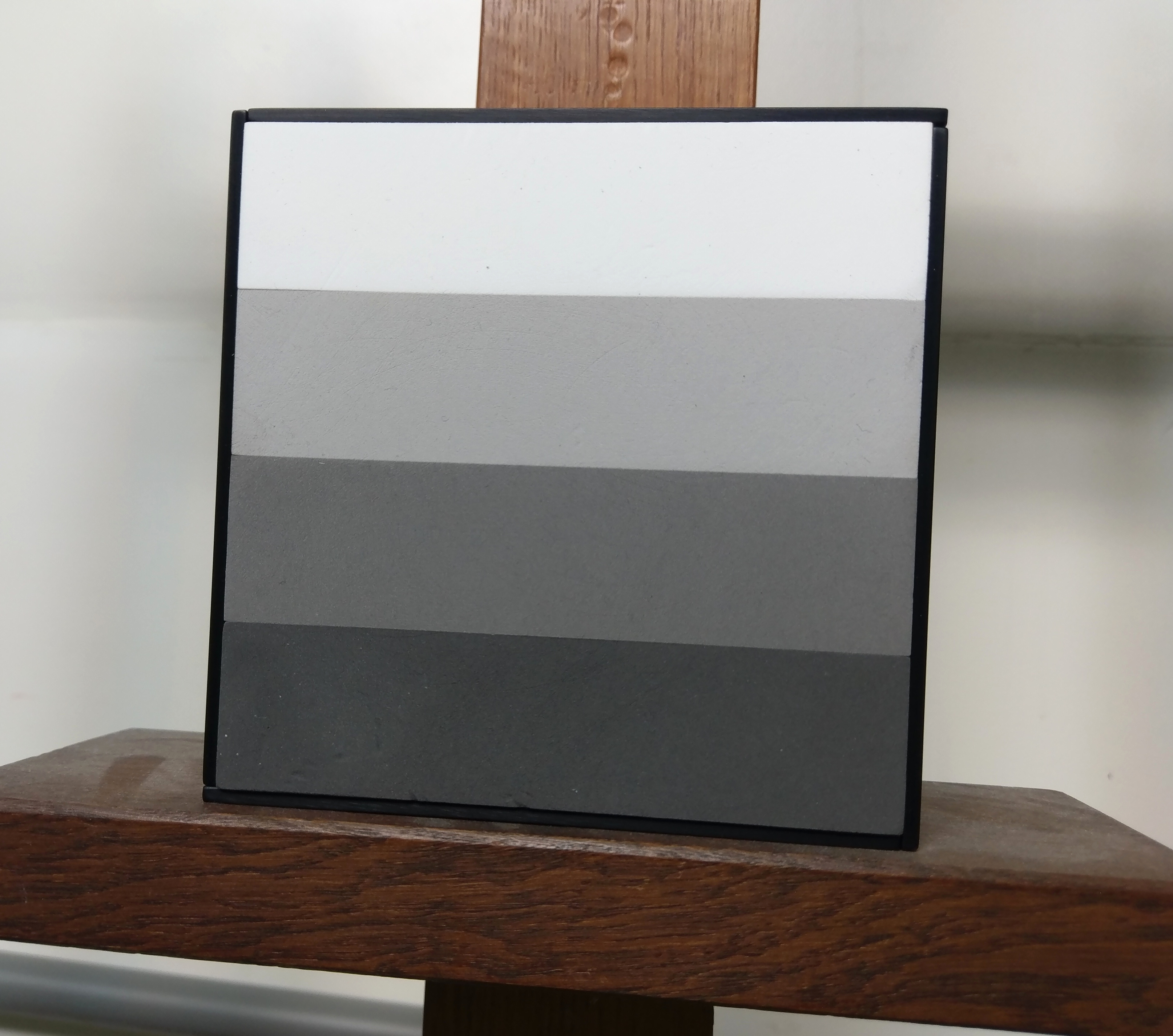}
\caption{Multi-step reflectance standard with reflectances at 99\%, 50\%, 25\% and 12.5\%}
\label{fig:multistep_spectralon}
\end{figure}

However, the quality of the radiometric calibration can often be improved by acquiring a set of different reference targets with a range of reflectivities. An example of such a target is shown in figure \ref{fig:multistep_spectralon}. Again, these must be acquired using the same parameters as those for the painting acquisition either during the same scan sequence or separately.

For many works of art, tuning the acquisition parameters in order to acquire a good image of the 99\% reference standard will result in sub-optimal results for the work of art itself. The maximum reflectance found in even a bright painting is rarely close to that of a 99\% reference target and dark paintings can have maximum reflectances of 50\% or less, meaning that half of the dynamic range of the hyperspectral sensor is lost. Indeed as detectors become non-linear when operated at close to saturation, users are recommended by the manufacturers to set their acquisition parameters in order to capture data below the saturation level and often at around 80\% or less. Thus, in practice, a dark painting would only use 40\% of the sensor's full dynamic range, significantly degrading the potential performance of the system.

By using several reference targets together, however, the integration time of the system can be set according to the requirements of the artwork itself rather than be limited by any particular reference target. The integration time of the acquisition can, therefore, be increased for darker works of art and set at an optimal level for them. In practice, this means performing a test scan over the brightest part of the work of art and increasing the integration time until the detector is at, for example, 80\% saturation. In this way, the acquisition uses as much of the available dynamic range as possible, significantly improving signal-to-noise and the quality of the resulting data.

An additional advantage to using multiple reference targets is the ability to combine the measurements from all of the (unsaturated) reference targets and calculate a more accurate linear fit, especially when using low light levels or with less sensitive or noisy equipment \citep{burger_hyperspectral_2005}.

If the integration time has been tuned to the work of art and only a single reference target is to be used, it should have a reflectance that is similar or less than the brightest part of the work of art and it is essential that the data from this target is not saturated.

A spectral image cube of one or more reference standards is critical for calibrating to reflectance factor and it is important that this data is of high quality and, in particular, that only those not saturated in any spectral band are used for radiometric calibration.

\subsubsection{Geometric Targets}

Push-broom hyperspectral systems are essentially line-scanners that acquire one single spatial line at a time. In order to acquire a full 2D image, it is necessary to physically move the camera perpendicular to this scan line along the Y-axis. In order to avoid errors in the aspect ratio of the resulting image, it is important to correctly calculate the scan speed along this direction of motion. Some hyperspectral camera manufacturers supply acquisition software that automatically calculates this scan speed, but many do not, thereby requiring the user to input a scan speed manually. This can often be tricky as tiny errors in the calculation of the scan speed can result in large accumulated errors over extended scan areas. Indeed, incorrect aspect ratios due to incorrectly calculated scan speeds were a common error seen in the RRT.

In order to better calculate the scan speed and to help correct for any residual errors during post-processing, a geometric calibration target can be placed in the scene. Such a geometric calibration target can be something as simple as a ruler or printed grid, which can be used to calculate the appropriate scan speed or can be used to re-scale the data during post-processing.

\subsection{Focusing and Image Sharpness}
\label{sec:focusing}

The alignment of the work of art is an important initial step in most hyperspectral imaging workflows. The work of art must be at an appropriate distance to the hyperspectral camera and as perpendicular as possible to both the camera and to the direction of scanning. This alignment is important for obtaining a good sharp focus across the entire field of view and to avoid introducing geometric errors.

Hyperspectral imaging systems need to capture as much light as possible and so typically use lenses with low F-numbers and, therefore, very open apertures. This, however, results in relatively shallow depths of field. Hyperspectral imaging was originally devised for use in remote sensing where focus is obtained at infinity and depth of field is less of an issue. For close-range imaging, however, obtaining a spectral image cube that is sharp and in-focus across the whole spectral range can be difficult. This is especially true for targets such as panel paintings which can often be warped or have uneven surfaces and sub-optimal focusing was indeed an issue for several participants in the RRT, where data were often blurred.



\begin{figure*}
\centering
\subfigure[Image sharpness with respect to focus position for each spectral band.]{%
  \label{fig:focus_metrics}
  \resizebox*{7.5cm}{!}{\includegraphics{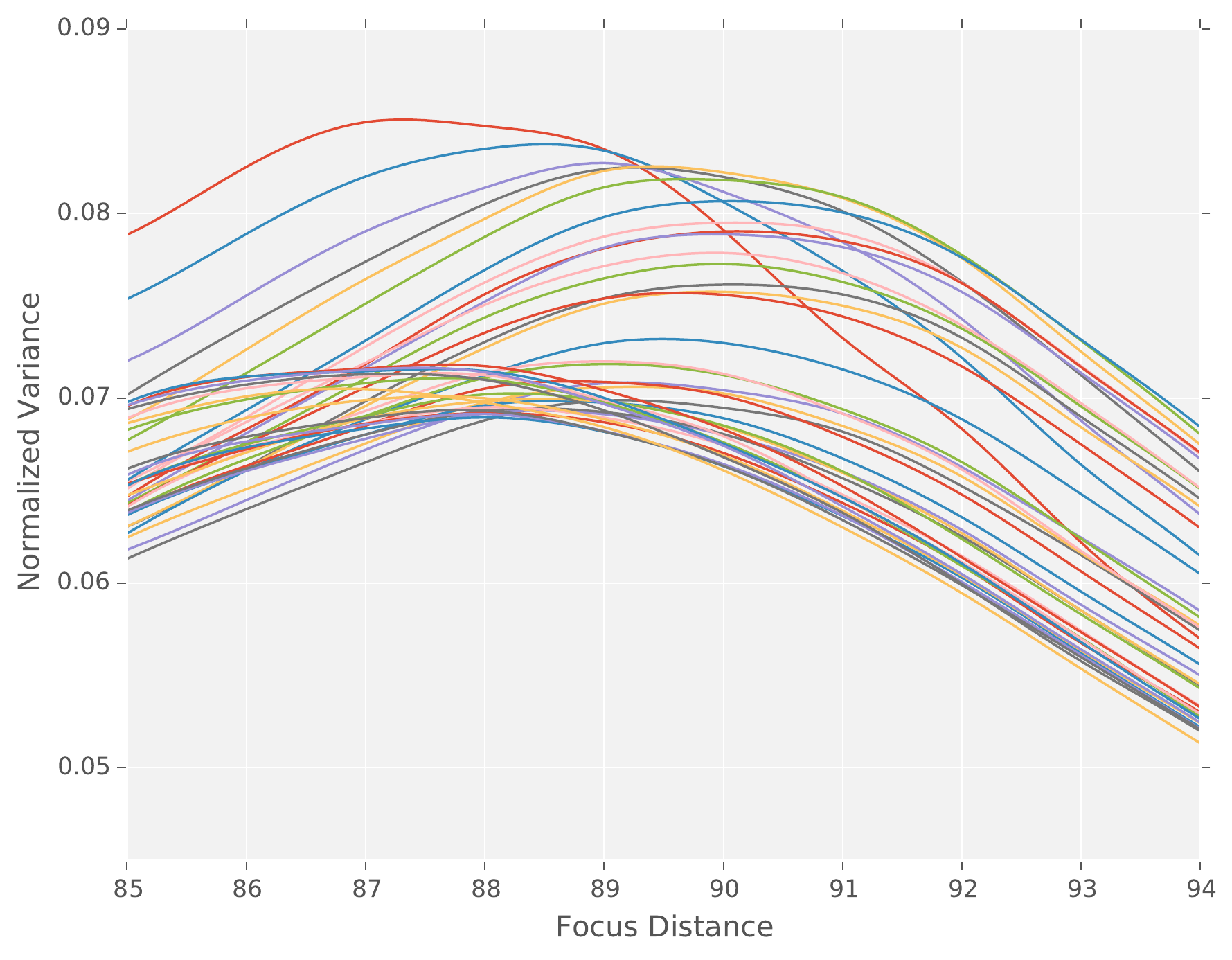}}}\hspace{5pt}
\subfigure[Optimal sharpness location as a function of wavelength.]{%
  \label{fig:focus_wavelengths}
  \resizebox*{7.5cm}{!}{\includegraphics{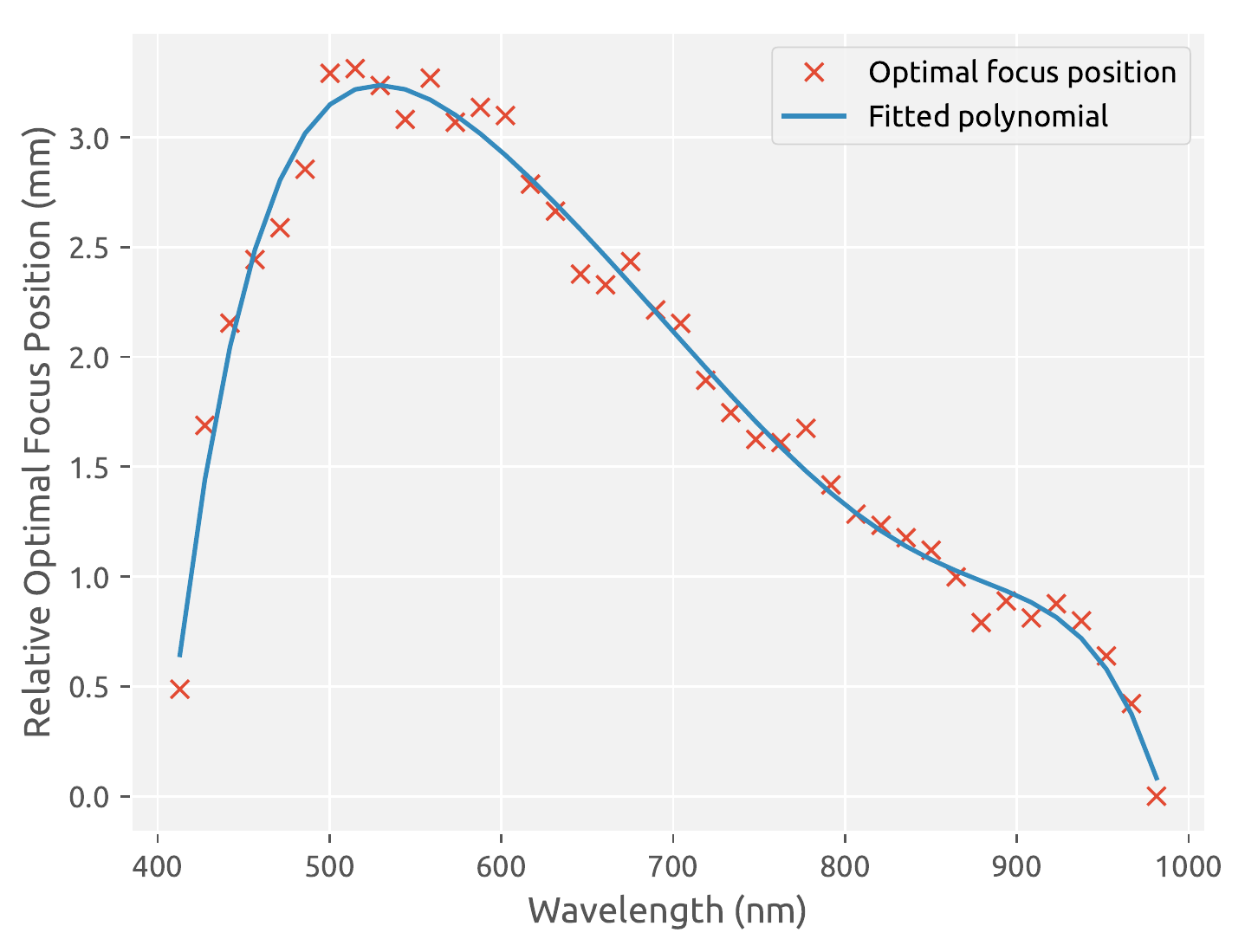}}}
\caption{Focus metrics for the HySpex VNIR1600 showing how the optimal focus distance varies with respect to wavelength.}
\label{fig:focus}
\end{figure*}

Hyperspectral acquisition systems can be either fixed-distance focusing or can be focused through the lens. In both cases, the need for appropriate focusing is essential. The ideal focus setting generally varies with respect to wavelength. Although spectral acquisition systems are designed to minimize spectral aberration, this is difficult to eliminate entirely. Finding the ideal focus is, therefore, a compromise over the entire spectral range. In figure \ref{fig:focus_metrics}, we see the results from measuring the level of sharpness of an image for each captured wavelength at different distances for a visible \textendash{} near infrared (VNIR) hyperspectral system with a fixed-distance lens. In this example, the measure of ``sharpness'' is the simple, but widely used variance \citep{groen_comparison_1985} of the image for each individual band. 

We see from figure \ref{fig:focus_metrics} that the peak for each wavelength occurs at a different distance of the camera to the target. If we plot the locations of the peaks with respect to wavelength, figure \ref{fig:focus_wavelengths}, we see that the ideal focus distance is different at each wavelength. Knowing this, it is possible to choose a focus setting either suited to the spectral region most of interest or which is a suitable compromise that provides good focus for the data as a whole.

An improvement to using a single focus distance would be to extend the depth of field through image processing techniques such as focus-stacking \citep{pieper_image_1983} 
 whereby multiple images are acquired at different focus distances and combined into a single image in focus at every wavelength.

\subsection{Filters}

\subsubsection{Equalization Filters}

\begin{figure*}
\centering
\subfigure[Quantum efficiency of Kodak KAI-2020 CCD.]{%
  \label{fig:quantum_efficiency}
  \resizebox*{7cm}{!}{\includegraphics{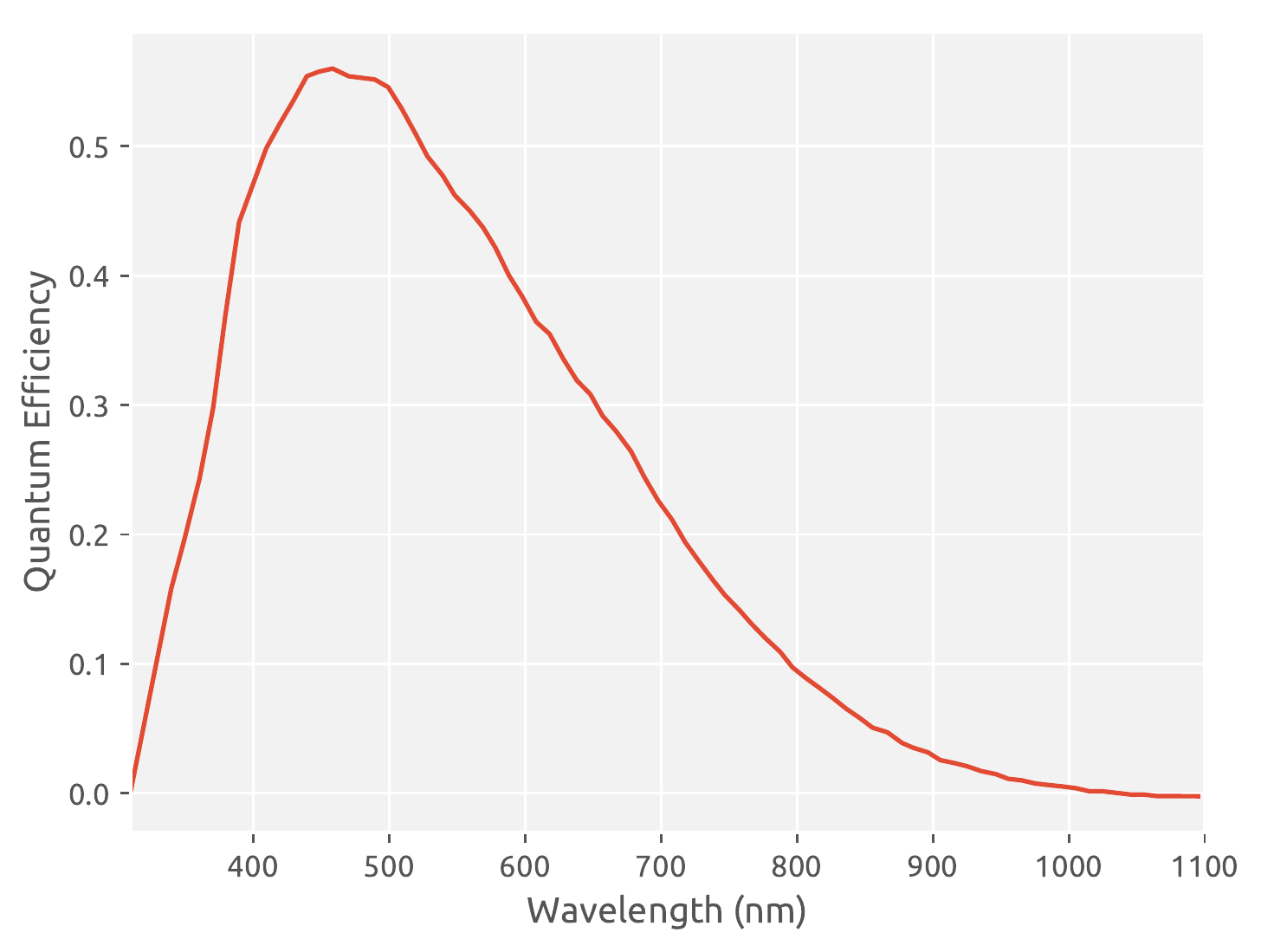}}}\hspace{5pt}
\subfigure[Spectral power distribution for EKE-ER 150W full spectrum lamp (Illumination Technologies Inc.).]{%
    \label{fig:illuminant}
  \resizebox*{7cm}{!}{\includegraphics{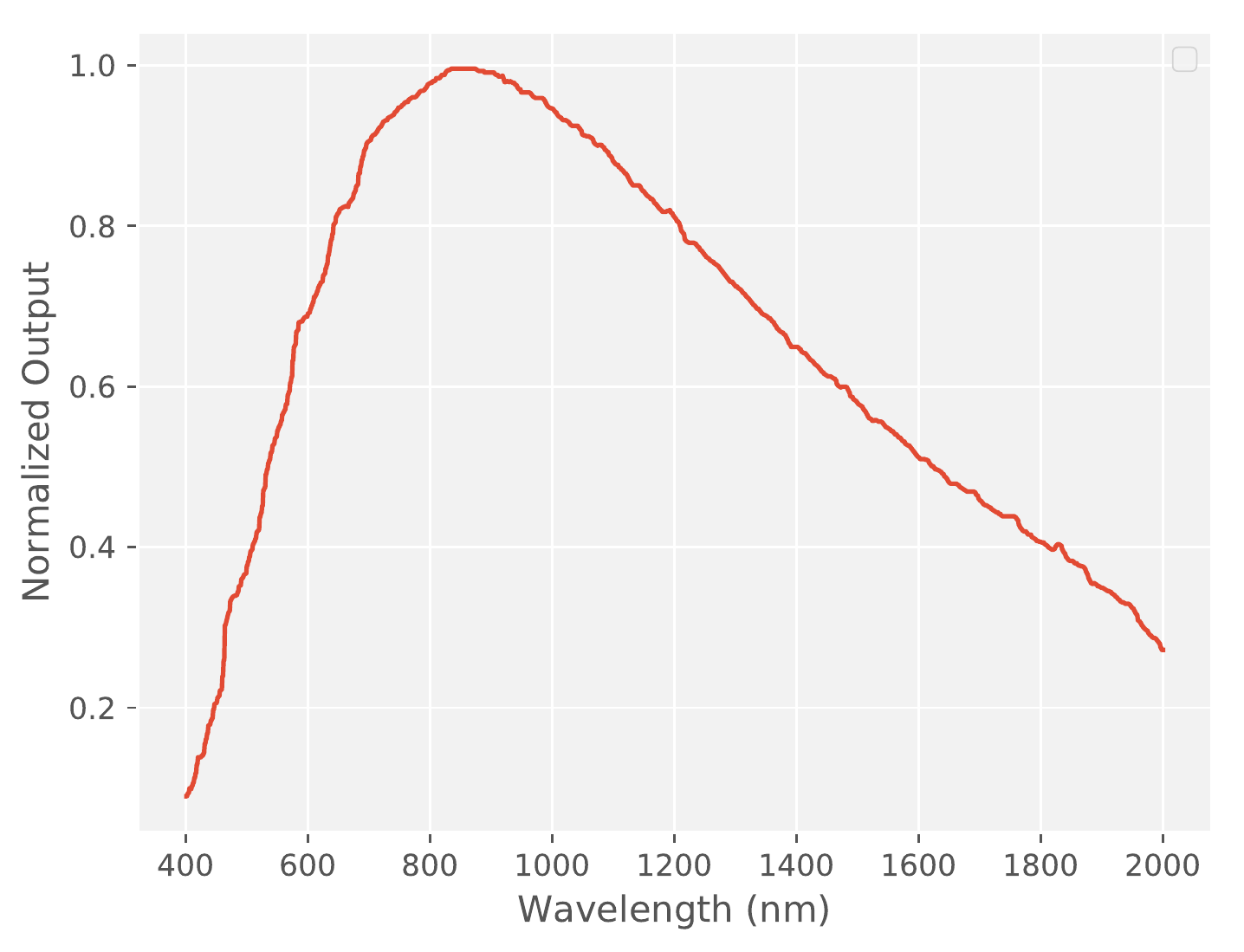}}}
\subfigure[Equalization filter transmission.]{%
    \label{fig:equalization_filter}
  \resizebox*{7cm}{!}{\includegraphics{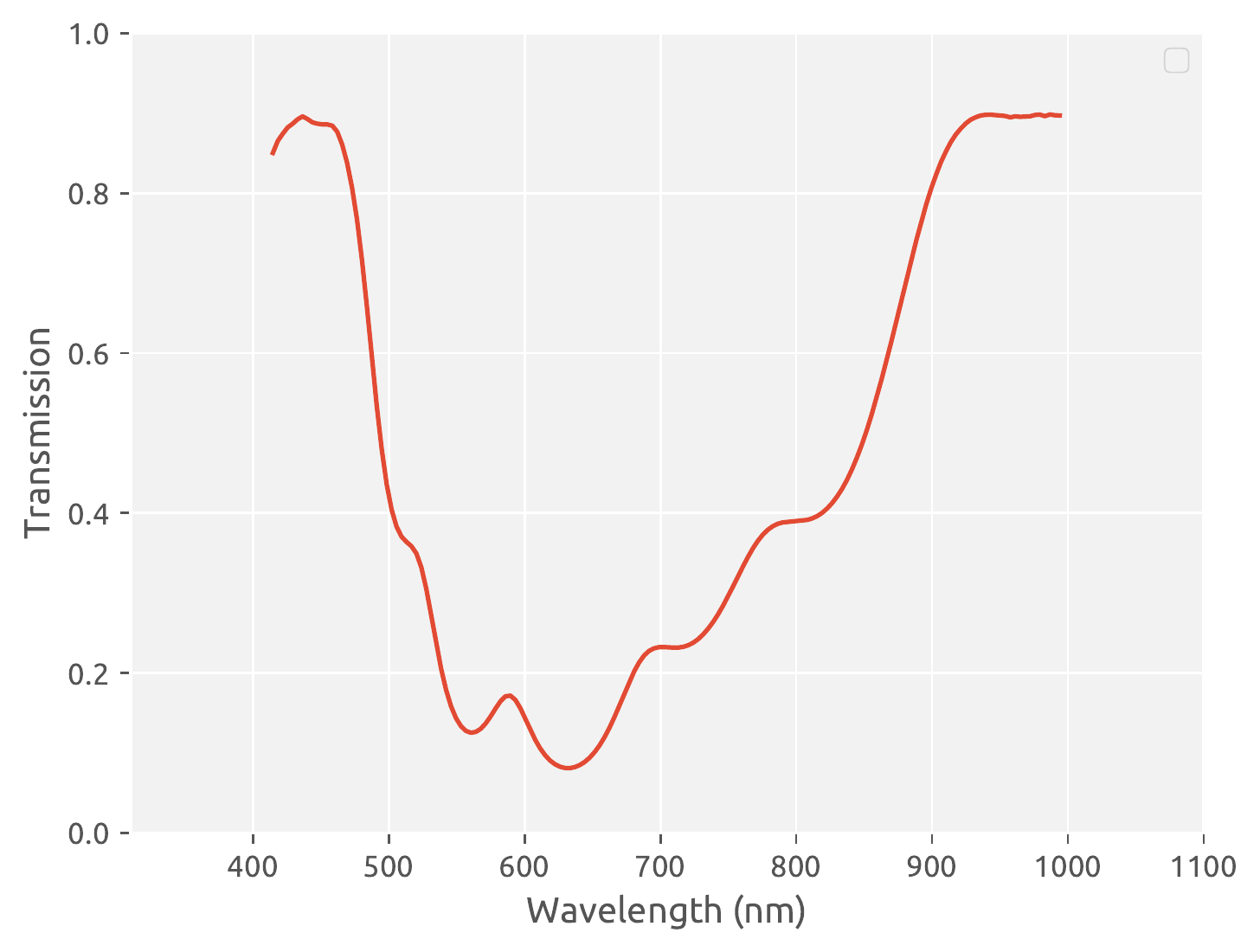}}}
\subfigure[Resultant spectral efficiencies after taking into account detector efficiency, illuminant power distribution and effect of an equalization filter.]{%
  \label{fig:system_efficiency}
  \resizebox*{7cm}{!}{\includegraphics{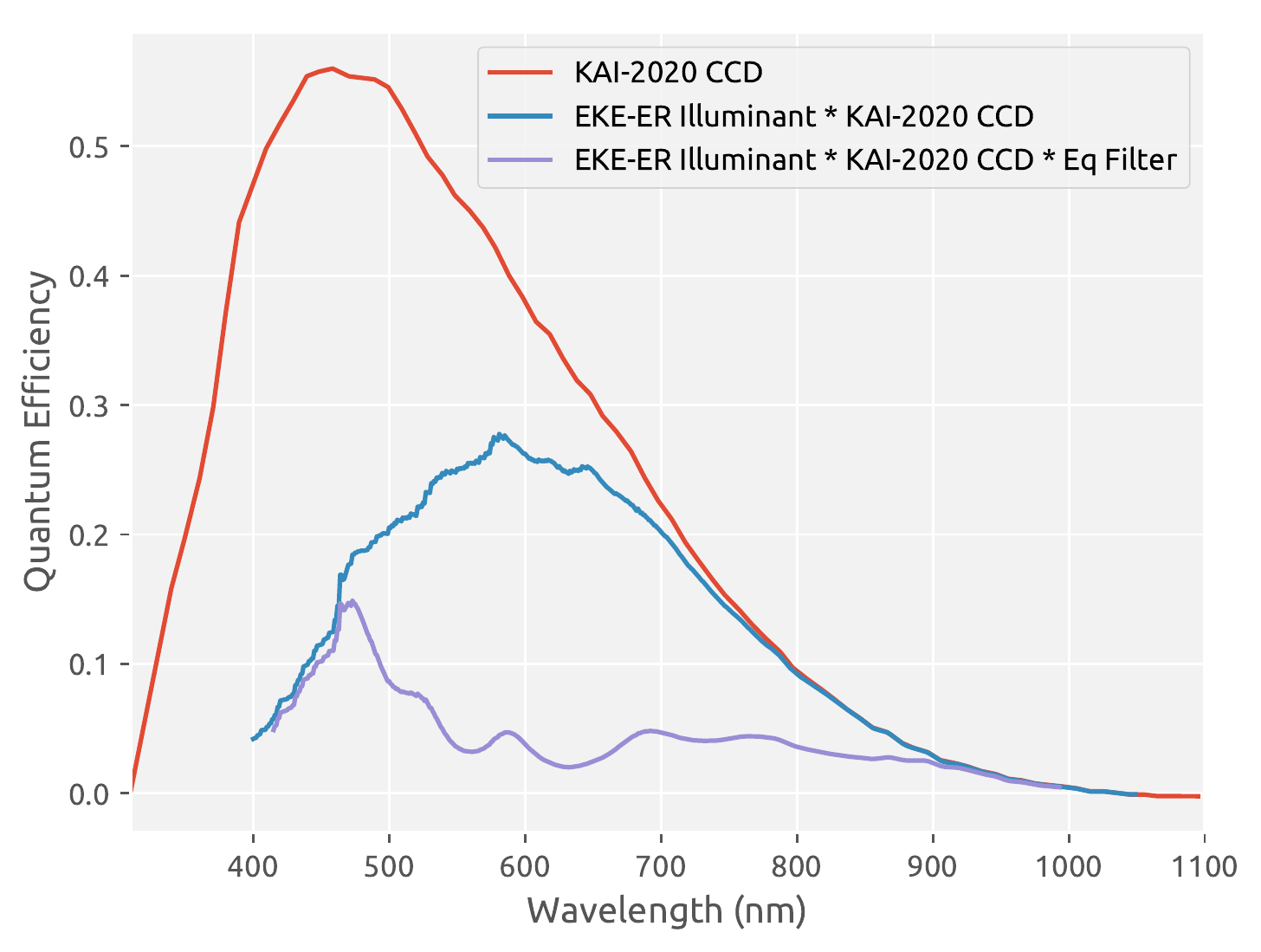}}}\hspace{5pt}
\caption{The quantum efficiency of the detector and the impact of the illuminant and equalization filter on it.}
\end{figure*}

Hyperspectral camera detectors use various semiconductors that have different optimal operating ranges and sensitivities. This sensitivity (or quantum efficiency) is dependent on the wavelength of the light being measured and is different for each semiconductor material. The factory-measured quantum efficiency curve from the Silicon-based CCD used in several VNIR hyperspectral cameras, such as the HySpex VNIR1600\footnote{manufactured by Norsk Elektro Optikk AS} is given in figure \ref{fig:quantum_efficiency}. The signal at the low and high extremes of the spectral range of the detector has a significantly lower sensitivity and is, therefore, prone to significant noise.

However, the sensitivity of the system as a whole is dependent not only on the camera detector, but also on the illumination used. The spectral power distribution of a typical extended spectral range illuminant for VNIR systems is given as a function of wavelength in figure \ref{fig:illuminant}, where we can see that although power at the upper limit of the detector's range towards 1000nm is excellent, the power towards the lower end at 400nm drops to very low levels. If we multiply the quantum efficiency with the output of the illuminant, we obtain the combined efficiency of the system (the second plot in figure \ref{fig:system_efficiency}) which shows that our peak sensitivity occurs at just under 600nm and that our sensitivity at 1000nm, despite the strong output of the light source at that wavelength, is very weak indeed with an efficiency that is over 50 times less than that at 600nm.

Unlike in a snapshot system (as typically used for multispectral imaging), where each spectral band is acquired independently, the integration time for a push-broom hyperspectral camera must be set globally for the entire spectral range, making data quality very variable between regions of high and low sensitivity. There are, nevertheless, several ways to improve the signal-to-noise at the extremes of the spectral range.

The first is to increase the intensities of the spectral content of the light sources in those wavelengths where our illuminant is weak and the detector has low sensitivity. Thus, in our example, this means in the blue region towards 400nm, where secondary narrow-wavelength lamps may be employed.

Alternatively, or in addition to the previous method, it is possible to use an equalization filter. Equalization filters are designed to reduce the sensitivity in the central most sensitive region of the detector's spectral range, thereby improving the \textit{relative} signal-to-noise at the extremes. This can greatly improve the overall signal-to-noise of the data, but necessarily requires longer integration times and reduces signal-to-noise for the central wavelengths.

The ideal equalization filter would be the exact inverse of the quantum efficiency. However, obtaining an exact inverse is difficult to obtain in practice and figure \ref{fig:equalization_filter} shows the measured transmission of a commercially available equalization filter. The combination of the illuminant, camera quantum efficiency and equalization filter is shown by the third plot in figure \ref{fig:system_efficiency}, which has a relatively flatter response over the spectral range, thereby improving the relative efficiency at 1000nm with respect to the average and maximum efficiencies. As we can see from the plot, however, the overall efficiency is much lower, therefore, requiring an integration time in this case around three times as long as it would be without a filter.

\subsubsection{Polarizing Filters}

For works of art that have areas that are very reflective and which have large amounts of specular reflection, it is possible to use polarizing filters in order to reduce this effect. Both the light source and hyperspectral camera optics require polarized filters with the filter in front of the optics rotated with respect to the those on the illumination in order to block light directly reflected from the surface.

The use of polarized filters in this way, however, necessarily reduces the light incident on the camera detector by half and will, therefore, require longer integration times.

\subsection{Frame Averaging}

A complementary technique to the use of equalization filters is frame averaging, whereby a number of co-incident scans of the same line on the work of art are averaged together to improve signal-to-noise \citep{kohler_photographic_1963}. This technique improves signal-to-noise by a factor of $\sqrt{N}$, where $N$ is the number of scans used in the averaging. In general, push-broom hyperspectral systems achieve this by slowing down or momentarily halting the acquisition scan movement and scanning a single line multiple times. This, however, increases the overall acquisition time, so may not be practical in many cases.

By combining the use of both an equalization filter and averaging, significant improvements in data quality can be achieved at the expense of increased scan time.

\section{Data Processing - Calibration Workflow}
\label{sec:processing}

\begin{figure*}[hb]
\centering
\includegraphics[width=1\textwidth]{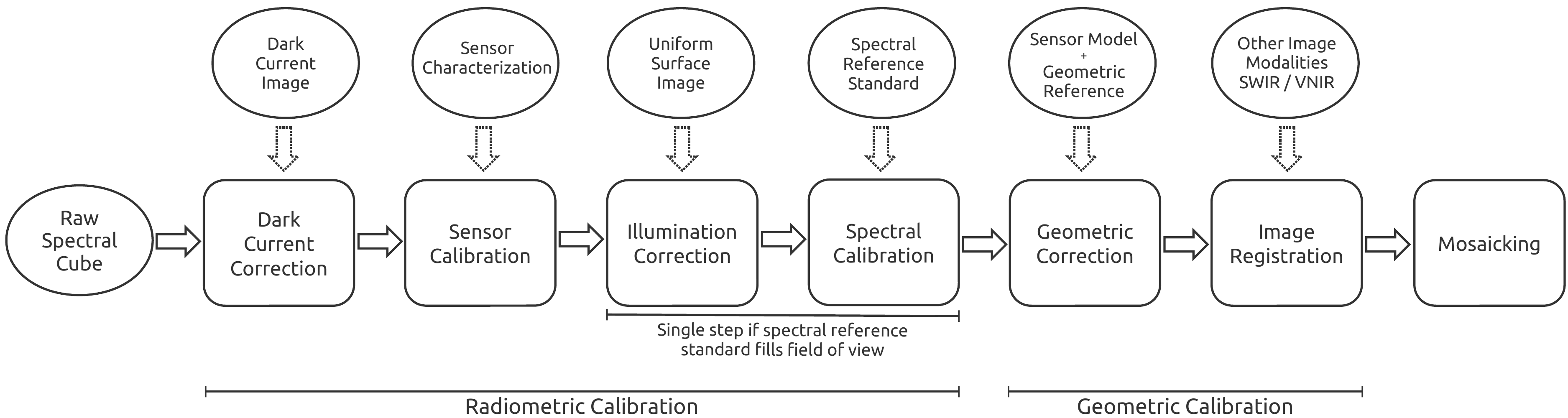}
\caption{Workflow for hyperspectral calibration showing the processing steps and the data necessary to perform them.}
\label{fig:calibration_workflow}
\end{figure*}

The acquired raw hyperspectral data will need to be processed to produce accurate and calibrated reflectance data. The processing can be divided into two domains: radiometric calibration and geometric calibration and an overview of the steps can be seen in figure \ref{fig:calibration_workflow}. As was seen in the RRT, all participants applied varying levels of radiometric calibration to their data. Geometric calibration, was, however, rarely carried out.

\subsection{Radiometric Calibration}
\label{sec:radiometric}

Radiometric calibration takes the raw pixel data and produces accurate quantitative reflectance values for each pixel. This step needs to compensate for the variations in the spectral sensitivity for each pixel and both the spatial inhomogeneity and the spectral content of the illumination. In addition, it must also take into account various sources of noise, which can often be significant in hyperspectral imaging.

The first step in the calibration workflow is to remove dark current (also sometimes referred to as fixed pattern) noise. Dark current is essentially electronic noise within the detector which increases with respect to integration time and the operating temperature of the camera. To correct for this, a dark current scan is acquired using the same acquisition settings as for the main scan with no light incident on the detector. In practice this involves blocking the camera optics with either a lens cap or by using an electronic shutter. This should be carried out several times in order to average the data, thereby minimizing random noise and revealing the underlying fixed pattern offset. For push-broom hyperspectral systems, this involves performing a scan over typically a hundred lines with the optics blocked. The dark current is then calculated by simply averaging the individual pixel values from each scan line to provide a dark current for each pixel on both the spatial and spectral axes. This dark current offset should then be subtracted pixel-wise from all subsequent scan data. Certain manufacturers include this functionality within their acquisition software, thereby automating and simplifying this step.

The following step involves correcting for the variability in the sensitivities of each pixel. This is a particular problem for MCT (Mercury Cadmium Telluride) SWIR detectors, which often exhibit very marked striping in their raw data \citep{rogalski_hgcdte_2005}. 
To correct this, an image must be acquired under constant and extremely uniform lighting conditions, such as in an integrating sphere. This characterization data is usually acquired by the manufacturer and often automatically corrected for by the acquisition software.

Spatial variability in the illumination and from the optics then also needs to be corrected for. This can be performed by acquiring a diffuse uniform grey or white target that fills the entire field of view and the data must be scaled accordingly.

Once, the pixel sensitivities and spatial illumination has been accounted for, a final spectral correction must be performed by acquiring an image of a known spectral reflectance standard. The spectral image cube can then be scaled to true reflectance factor.

For push-broom hyperspectral systems, it is often practical to combine these last 2 steps into a single step by using a spectral reflectance standard that is large enough to fill the entire field of view.

\subsubsection{Reflectance Standard Reliability}

The use of one or more reflectance standards is essential in order to calibrate spectral reflectance data. Often, however, this is carried out in a very rudimentary form where the spectral image cube is simply scaled to the nominal reflectance factor given for the reference target - in practice this means dividing the spectral image cube by the measured pixel values of an image of a reflectance standard. Indeed the vast majority of the participants of the RRT used a reference target in this way.

\begin{figure*}[ht]
\centering
\subfigure[Spectral reflectance for three different 99\% targets.]{%
    \label{fig:spectralon_reflectance_99}
  \resizebox*{7.5cm}{!}{\includegraphics{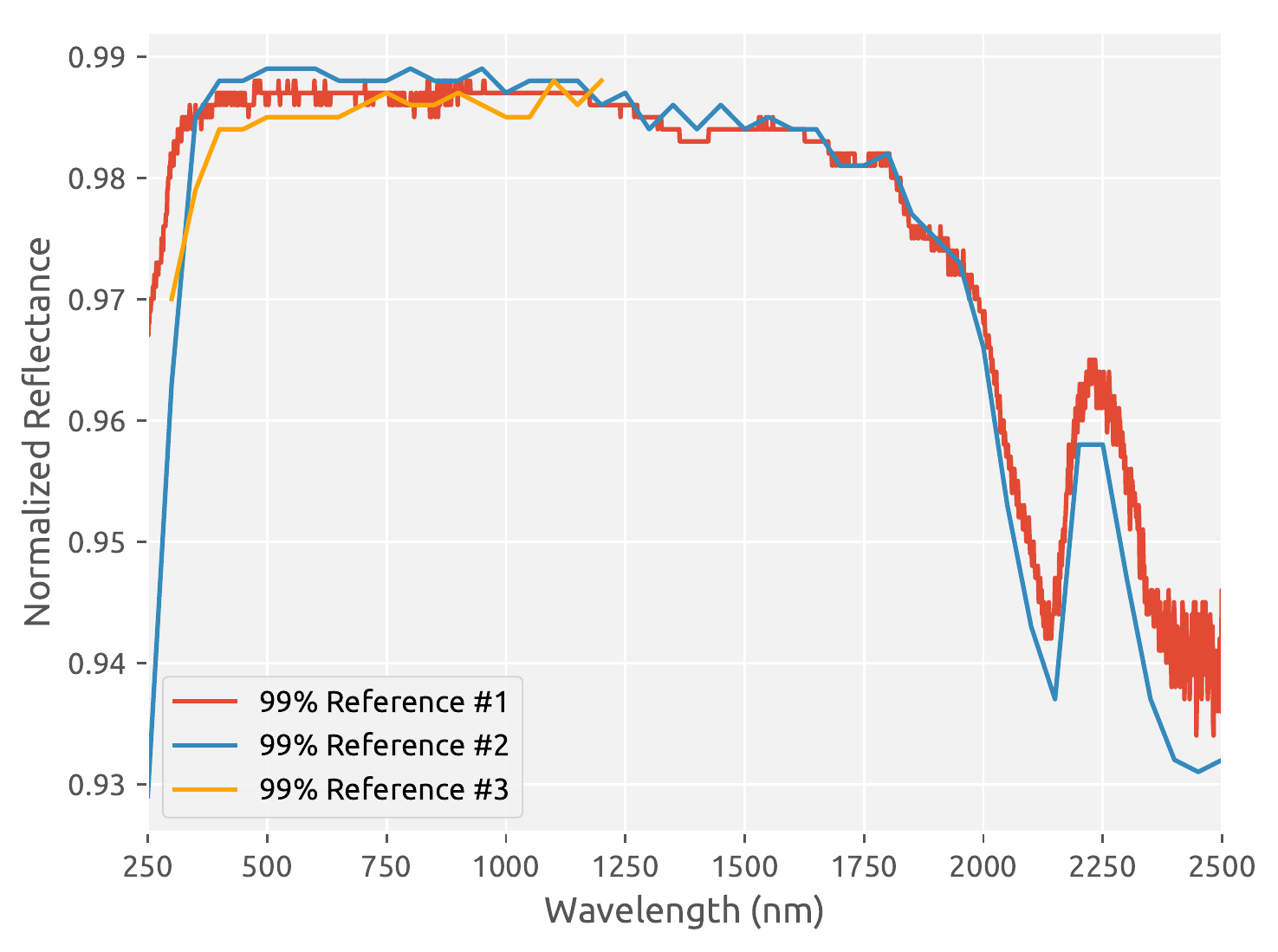}}}\hspace{5pt}
\subfigure[Spectral reflectance for two different 50\% targets.]{%
    \label{fig:spectralon_reflectance_50}
  \resizebox*{7.5cm}{!}{\includegraphics{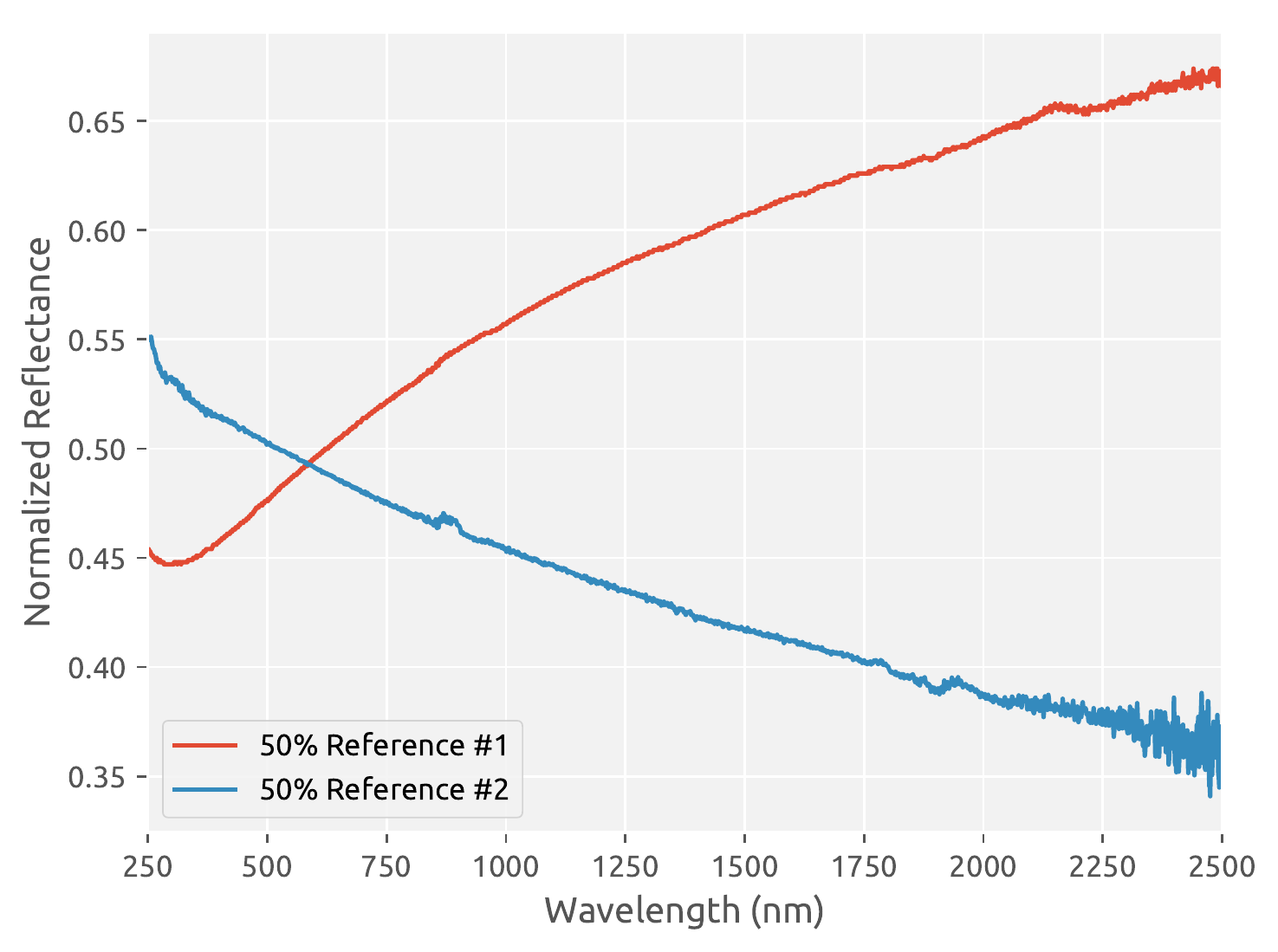}}}
\caption{Variability in spectral reflectance for various reflectance standards at 99\% and 50\% nominal reflectance.}
\label{fig:spectralon_reflectance}
\end{figure*}

These spectral reflectance standards are widely considered to have a constant reflectance at their nominal reflectance value. However, although they are highly uniform compared to other reflective materials, their reflectances are never perfectly constant and can vary with respect to wavelength. Reflectance standards are usually supplied with individual traceable laboratory-certified absolute reference reflectance values across the spectral range for which these targets are intended to be used, typically covering the UV, visible and SWIR (short-wave infrared) spectral regions. Figure \ref{fig:spectralon_reflectance} shows the variability that exists for the certified reflectances of five different reflectance standards with nominal reflectances of 99\% and 50\%.

Although the 99\% reflectance targets are close to uniform over the visible to near infrared region, this is no longer the case from 1500nm to 2500nm. Indeed there are inflections in the spectral curves at around 2130 and 2230nm for all three 99\% targets with the reflectance in reality ranging from 90\% to 99\%. For the 50\% reflectance targets examined, the differences are greater still with our two examples exhibiting diverging gradients and real reflectances which vary from 37\% to 67\%.

Thus, in order to accurately calibrate hyperspectral data to absolute reflectance factor, especially in the short-wave infrared region, the supplied certified reference reflectance values should be used to normalize the hyperspectral data \textit{individually} for each spectral band rather than just assuming a constant fixed reflectance across the entire spectral range.

\subsubsection{Reflectance Standard Statistics}

There are also a number of subtle statistical errors that can be introduced when processing reflectance standards, as even the best maintained reference standards can get dirty, causing incorrect values to be obtained. One way to avoid this is to exploit the fact that we have a line-scan set-up and can average the data points for each column of the reference target. In this way, the effect of dirt can be reduced. However, such a method, although an improvement on a spot measurement is still subject to bias if done naively. In order to better understand our data, it is useful to first analyze the results obtained from the reference targets in terms of homogeneity and spatial behavior.

\begin{figure*}
\centering
\subfigure[Raw spatial line profile across a 50\% reflectance target of an acquired band at 600nm.]{%
  \label{fig:spectralon_profile}
  \resizebox*{7.5cm}{!}{\includegraphics{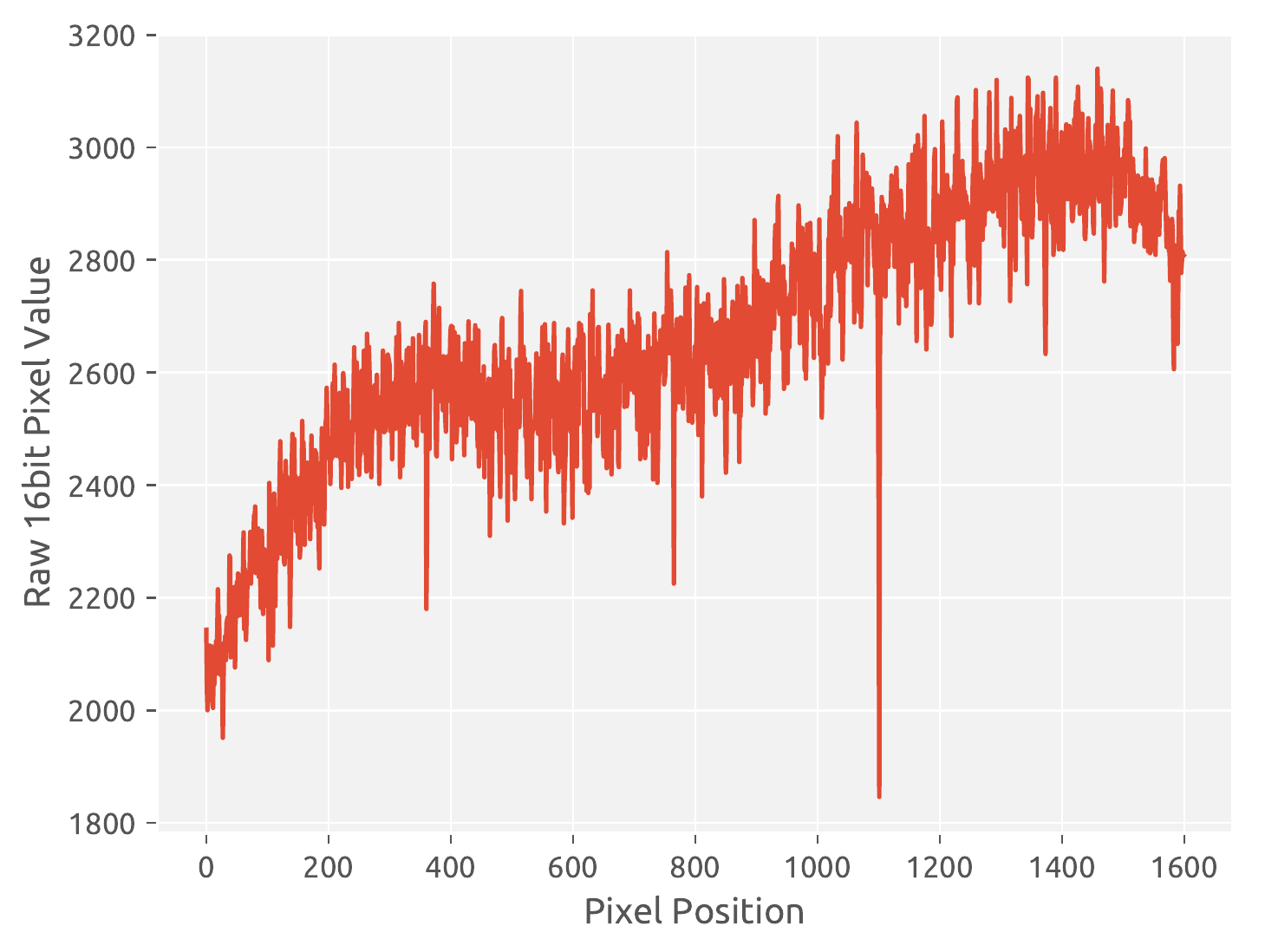}}}\hspace{5pt}
\subfigure[Histogram of raw pixel intensities of 50\% reference target for a single pixel along Y-axis.]{%
  \label{fig:spectralon_histogram}
  \resizebox*{7.5cm}{!}{\includegraphics{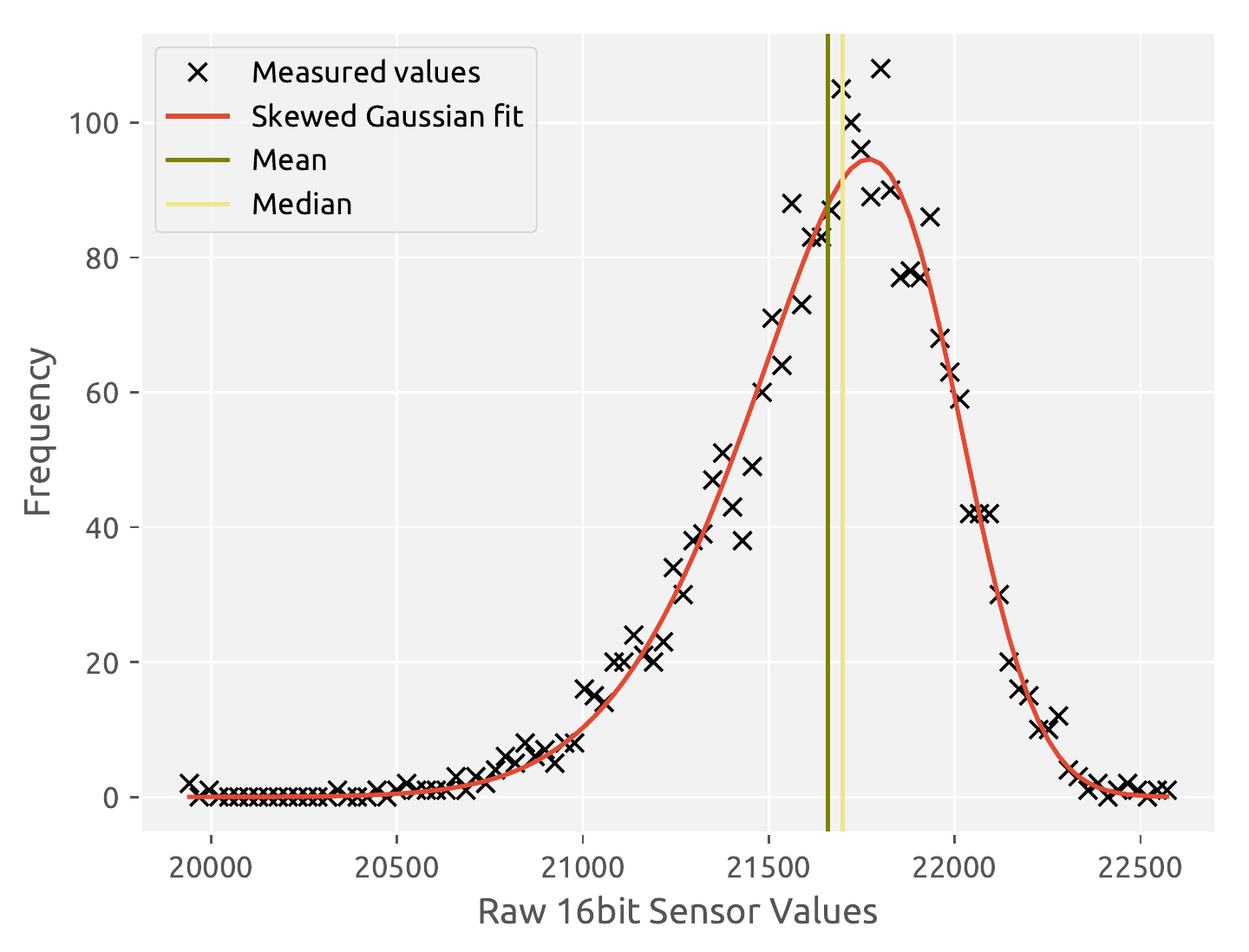}}}
\caption{The raw spatial line profile from a hyperspectral camera of a single target (left) and the histogram of pixel values for a single point on the profile when scanned along the Y-axis (right).}
\label{fig:spectralon_statistics}
\end{figure*}

A typical measured spatial profile at a particular wavelength from a line along the X (horizontal) direction of a reference target can be seen in figure \ref{fig:spectralon_profile}. The individual pixel values within the spatial profile are a function of both the reflectance at each point, the illumination and camera noise. Ideally this distribution should be as flat as possible to ensure that the signal-to-noise is constant across the field of view. However, a perfectly even distribution is difficult to achieve in practice and as we can see from the example, the measured values can vary quite markedly across the field of view despite careful adjustment of the light sources and of their alignment. The signal is also quite noisy with several outliers due to dirt or imperfections on the target. In order to reduce the effect of this noise, it is usual with push-broom hyperspectral systems to scan along the Y direction of the reference standard and average along this axis in order to obtain a relatively noise-free measure of the reflectance for each pixel.

However, although the average is the most widely used statistic, dirt and other imperfections have an asymmetric effect on the distribution resulting in a biased result. For a more accurate statistic, a skewed Gaussian distribution \citep{azzalini_multivariate_1996} provides a better fit and we see from figure \ref{fig:spectralon_histogram} the skewed form of the distribution of pixel values and the close fit provided by the skewed Gaussian model. Both the mean and median values understate the true value due to the assymetric nature of the outliers and, in this example, there is almost a 5\% difference in the calculated result, which will impact any spectral values calculated from this.

\subsection{Noise Reduction}

Hyperspectral systems are particularly prone to noise, given the narrow wavelengths used and the wide variability in quantum efficiency (as already seen in figure \ref{fig:quantum_efficiency}) across the spectral range of the sensor. Although an optimal use of illumination and integration time combined with the use of equalization filters and averaging can significantly improve the quality of the data, noise will invariably exist, particularly at the extremes of the spectral range of the sensor. And this noise can have a significant negative effect on the use of hyperspectral data for tasks such as classification, anomaly detection or materials identification \citep{landgrebe_noise_1986}.

Denoizing through post-processing is often, therefore, a necessary step in maximizing the utility of the data obtained. A wide range of noise reduction methods exist in image processing. Classic methods developed for greyscale or color images include linear filtering methods such as gaussian blurring, as well as a number of non-linear methods, such as median filtering.

However, spectral data consists of a sequence of highly correlated spectral bands, and it is advantageous to take this extra dimensionality into account. A considerable amount of research exists covering denoising methods adapted to hyperspectral data. 
These generally consist of adapting existing 2D image denoising methods to take into account the spectral dimension. Examples of these adapted methods include the use of wavelets \citep{rasti_hyperspectral_2012}, an extension to non-local means \citep{qian_3-d_2012}, variational methods \citep{mansouri_adaptive_2015} and combined spectral-spatial approaches \citep{yuan_hyperspectral_2012}.

\subsection{Geometric Calibration}
\label{sec:geometric}

Geometric calibration is rarely carried out for laboratory-based hyperspectral imaging as, in general, only the spectral values are of interest or the data is to be used in isolation. Such calibration is, therefore, seldom available in the acquisition or calibration software provided by the manufacturer. However, if registration to other imaging modalities is intended or if mosaicking will be carried out, geometric calibration \citep{spiclin_geometric_2010, khan_hytexila_2018} will need to be performed on the acquired data.

Hyperspectral systems produce geometric distortions due to the scan motion, to the arrangement of dispersion optics, from the lens itself and also from alignment errors with respect to the work of art. The RRT results showed that not only were the spectral results highly variable, but the resulting image geometries were often equally so \citep{macdonald_assessment_2017}.

\begin{figure}[ht]
\centering
\includegraphics[width=\linewidth]{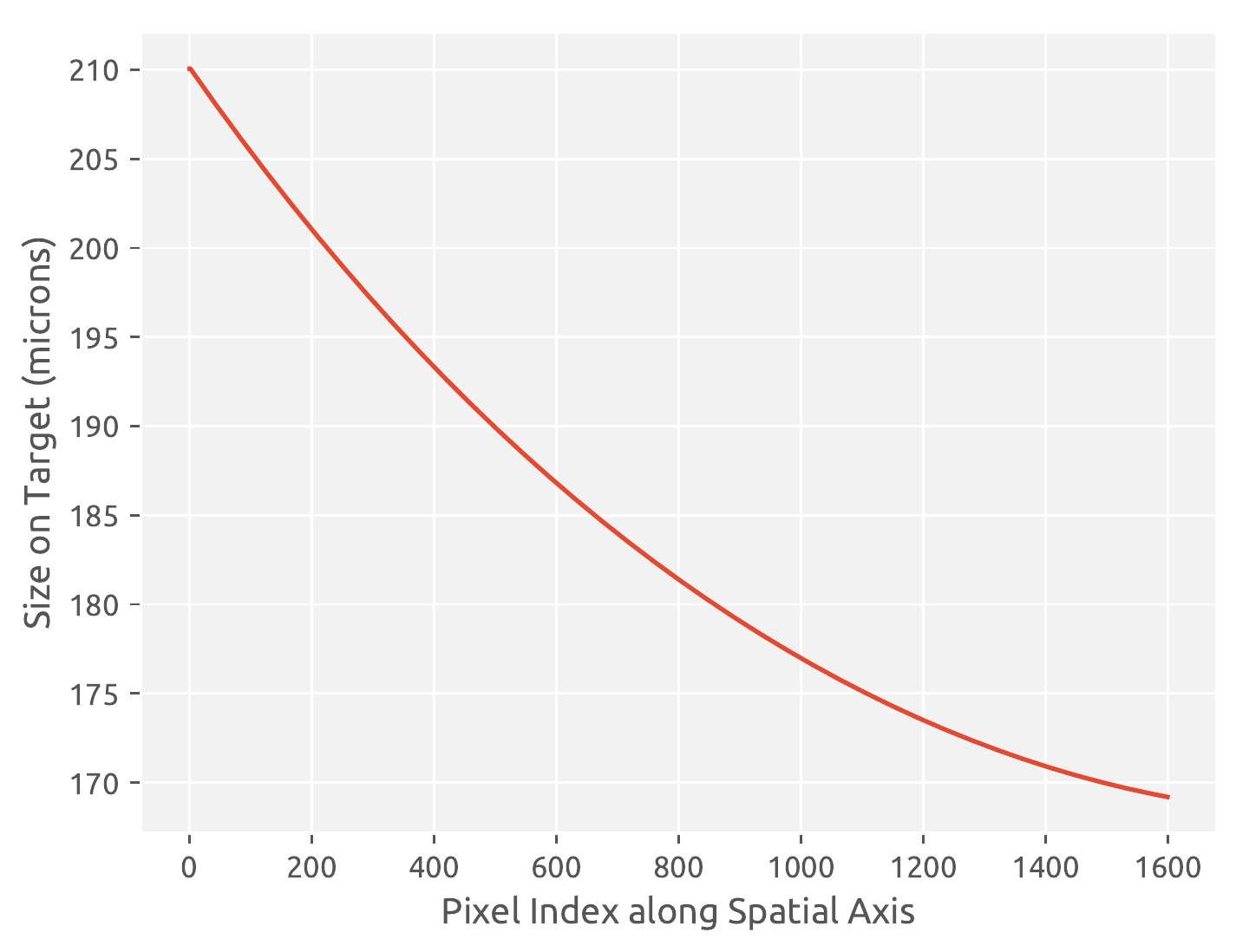}
\caption{Variation in effective pixel size for each pixel along X-axis for the HySpex VNIR1600}
\label{fig:sensor_model}
\end{figure}

The complex multi-component optics often used in hyperspectral systems can produce non-parametric distortions that require the characterization of a sensor model in order to correct. A precise sensor model can be produced by using a micrometer and measuring the angle of view for each pixel and several system manufacturers provide such characterization data. A typical sensor model is given in figure \ref{fig:sensor_model} and we see that there are differences of up to 20\% in the effective pixel sizes across the field of view that must be corrected if spatially accurate data is to be obtained.

In addition, if the scan speed has not been correctly calculated, the aspect ratio of the resulting spectral image cube may be incorrect. This can be the case even if the correct scan speed has been calculated due to tiny residual errors in the calculation or in the precision of the movement of Y-axis translation stage, which can accumulate over extended scan areas. In order to correct for this, the geometric calibration target should be used to re-scale the data appropriately along one of the axes.


\subsection{Spatial Co-registration}

It is often the case that data from spectral cameras with different spectral ranges are available, such as VNIR and SWIR. Data with an extended spectral range can be important for segmentation and materials identification. In order to make full use of this, it is necessary to co-register the data cubes into a single spectral image cube. SWIR detectors, such as those based on MCT or InGaS (Indium Gallium Arsenide) have much lower numbers of pixels (typically 256-320 pixels for MCT and up to 640 pixels for InGaAs), than VNIR cameras and so this step necessarily involves the resampling of data.

Although several participants of the RRT acquired both VNIR and SWIR spectral data, none were able to provide combined co-registered data cubes. Indeed, these complementary data sets were treated as if they were independent of each other.

If geometric correction has been applied, co-registration should only require alignment and resampling in order to match the spatial pixel sampling. However, even after geometric calibration, small errors can still exist due to minor optical aberrations etc. Co-registration can be optimized by several methods, such as, for example, feature-based matching over an overlapping spectral range \citep{schwind_improving_2014}, which can provide an automated and robust result with sub-pixel level accuracy.


\subsection{Mosaicking}
\label{sec:mosaicking}

Hyperspectral detectors have made great progress in terms of both spectral and spatial resolution. However, spatial resolutions remain limited to at best 2200 pixels for current generation VNIR cameras. Thus, in order to obtain sufficient spatial resolution of a work of art, it is necessary to perform mosaicking of the data cube as the final step in the processing pipeline.

Such mosaicking is now performed routinely for remote sensing data \citep{palubinskas_mosaicking_2003}, which often has the benefit of both accurate GPS positional and height data to help align and ortho-rectify hyperspectral scans. Laboratory push-broom hyperspectral systems that possess both X- and Y-axes are able to scan very regularly spaced strips. However, this level of precision is often insufficient and any variability in the planarity of the artwork can result in distortions due to the close-range nature of the acquisition. Many panel paintings, for example, display warping and even canvas paintings are rarely perfectly flat due to the chassis or from the topographical relief produced by uneven or thick paint layers.

Although the majority of participants in the RRT carried out mosaicking manually using software such as ENVI\footnote{Harris Geospatial Solutions, Inc.}, more advanced methods for both mosaicking and registration have been used in the cultural heritage field \citep{conover_automatic_2015}.

\section{Software}

A range of both commercial and open-source software exists for handling data from hyperspectral imaging systems, which can be used to perform a calibration workflow. These include commercial remote sensing software such as ENVI, open source software such as SpectralPython\footnote{\url{https://www.spectralpython.net/}} and Orfeo Toolbox\footnote{\url{https://www.orfeo-toolbox.org/}} as well as the manufacturer supplied acquisition software which can occasionally include basic calibration functionality. However, this software is often ill-suited and limited for use on works of art, lacking in key functionality or difficult to use. Indeed, commercial spectral imaging software can be prohibitively expensive for most cultural heritage institutions.

The participants of the RRT used a mixture of commercial remote sensing software such as ENVI, custom-made tools and generic data processing packages such as Matlab\footnote{Mathworks Inc.}. Many participants had difficulties in using the software and cited the lack of specific functionality adapted to the imaging of works of art.

Therefore, in order to address this issue and allow cultural heritage users to more easily process their data, a suite of custom open-source software tools has been developed and made available\footnote{\url{https://hyperspectral-calibration.github.io}} that provides efficient implementations of the various steps of the calibration workflow described in this paper. By developing software collaboratively within the cultural heritage community, it will be possible to extend the software to handle the variety of equipment that is used and provide software adapted to the various needs found within cultural heritage. The use of a common set of software specifically designed for the cultural heritage sector will, furthermore, help improve the quality of hyperspectral data and foster the use of hyperspectral imaging more generally.

\section{Conclusion}

Hyperspectral imaging has become an increasingly used tool in the analysis of works of art. However, the quality of the acquired data and the processing of that data to produce accurate and reproducible spectral image cubes can be a challenge to many cultural heritage users. As seen in the COSCH RRT (round-robin test), a significant number of users had difficulties in acquiring high fidelity data 

The calibration of data that is both spectrally and spatially accurate is an essential step in order to obtain useful and relevant results from hyperspectral imaging. Data that is too noisy or inaccurate will produce sub-optimal results when used for pigment mapping, the detection of hidden features, change detection or for quantitative spectral documentation. To help address this, therefore, we have examined the specific acquisition and calibration workflows required for the hyperspectral imaging of works of art. These workflows include the key parameters that must be addressed during acquisition and the essential steps and issues at each of the steps required during post-processing in order to fully calibrate hyperspectral data.

To accompany this, we have also released a suite of open-source software tools that will enable users to more easily apply the steps described within the workflow. In addition, we have described a number of areas where image quality can be compromised and have proposed techniques to mitigate these problems and help improve overall data quality.

\section{Acknowledgments}

The authors would like to thank the participants in the spectral imaging round-robin test, which was carried out as part of the European Union COST Action TD1201 \textit{Colour and Space in Cultural Heritage} (COSCH)\footnote{\url{https://cosch.info}}.

\bibliographystyle{abbrvnat}
\bibliography{main}

\end{document}